\documentclass[letterpaper,10pt,oneside,conference,final]{IEEEtran}

\usepackage{graphicx}
\usepackage{algpseudocode}
\usepackage{algorithm}
\usepackage[english]{babel}
\usepackage{amssymb}
\usepackage{amsmath}
\usepackage{mathtools}
\usepackage{subfigure}
\usepackage[space]{cite}
\usepackage{xcolor}
\usepackage{bm}
\usepackage[utf8]{inputenc}

\algnewcommand{\LineComment}[1]{\State \(\triangleright\) #1}

\begin{document}


\IEEEoverridecommandlockouts

\title{Uplink Performance Evaluation of Massive MU-MIMO Systems}
  
\author{\authorblockN{Felipe A. P. de Figueiredo, Jo\~{a}o Paulo Miranda, Fabrício L. Figueiredo and Fabbryccio A. C. M. Cardoso
}
\authorblockA{CPqD--Research and Development Center on Telecommunications, Brazil. \\ Email: [felipep, jmiranda, fabricio, fcardoso]@cpqd.com.br}}
  
\maketitle
  
\begin{abstract}
The present paper deals with an OFDM-based uplink within a multi-user MIMO (MU-MIMO) system where a massive MIMO approach is employed. In this context, the linear detectors Minimum Mean-Squared Error (MMSE), Zero Forcing (ZF) and Maximum Ratio Combining (MRC) are considered and assessed. This papers includes Bit Error Rate (BER) results for uncoded QPSK/OFDM transmissions through a flat Rayleigh fading channel under the assumption of perfect power control and channel estimation. BER results are obtained through Monte Carlo simulations. Performance results are discussed in detail and we confirm the achievable "massive MIMO" effects, even for a reduced complexity detection technique, when the number of receive antennas at BS is much larger than the number of transmit antennas.
\end{abstract}


\section{Introduction}
MIMO technologies has been of utmost importance for the development and advance of broadband wireless communications systems in the last decades \cite{paulraj:mimo_overview}. By following and extending the ideas presented in \cite{foschini:layered_space_time_arch}, MIMO systems are now able to provide extremely high bandwidth efficiency and reliable transmissions at data rates that surpass 1 Gigabit/s. Suitable MIMO detection techniques offering both high performance and low complexity have been of extreme necessity for the technological improvements achieved in this area \cite{paulraj:mimo_overview, mietzner:survey, Larsson:lecture_notes}. The last decade has seen the successful introduction and implementation of MU-MIMO systems in several broadband communication standards \cite{gesbert:shifting_mimo}. 

MU-MIMO systems enhance the communication capabilities of its user terminals by applying space-division multiple access (SDMA) to allow multiple transmitters to send separate signals and multiple receivers to receive separate signals simultaneously in the same band. Therefore, in such systems, the more antennas at the Base Station (BS), the more users can simultaneously communicate in the same time-frequency resource \cite{hoydis:how_many_anetnnas}.

In very recent years, MU-MIMO systems have been adopting a very large number of antennas at the BS. By very large we mean that the number of antennas at the BS is much larger (in the order of at least 10$\times$) than the number of user terminal antennas within a cell. This idea was first proposed by Marzetta in \cite{marzetta:noncooperative}. This massive MIMO approach is recommendable due to the following characteristics: simple linear detection techniques become nearly optimal; both multi-user interference and small-scale fading effects tend to vanish; both power and spectral efficiency increase substantially \cite{hoydis:how_many_anetnnas,marzetta:noncooperative,rusek:scaling_up_mimo,Ngo:energy_in_mu_mimo}. 

The present paper considers the case of an OFDM-based uplink within a massive MU-MIMO system where the BS  is equipped with a large number of receive antennas (at least ten times greater than the number of transmit antennas) and adopts only simple, linear detection techniques. 

This paper is structured as follows. In Section II we introduce the uplink channel model adopted for this work. Section III presents some linear detection techniques, which can be employed in the uplink of massive MU-MIMO systems. In Section IV, this paper includes BER results for uncoded QPSK/OFDM transmissions through a flat Rayleigh fading channel where we assume perfect power control and channel estimation. The BER results are obtained by Monte Carlo simulations, which involves a bit error counting procedure. Finally, Section V gives some conclusions of the paper.

\section{Uplink Channel Model}
Massive MU-MIMO is a system where a base station (BS) equipped with a large number of antennas simultaneously serves several users in the same frequency band. Due to the large number of degrees-of-freedom available for each user, massive MU-MIMO can provide a very high data rate (due to a high multiplexing gain) and communication reliability with simple linear processing.

Consider a Massive MU-MIMO BS with $M$ receive antennas and that serves $K$ single-antenna users. Denote the channel coefficient from the $k$-th user to the $m$-th antenna of the BS as $h_{k,m}$, which is equal to a complex small-scale fading factor times an amplitude factor that accounts for geometric attenuation and large-scale fading:
\begin{equation}\label{eq:chan_coeffs}
h_{k,m} = g_{k,m} \sqrt{d_{k}},
\end{equation}
where $g_{k,m}$ and $d_{k}$ represent complex small and large scale fading coefficients, respectively. The small-scale fading coefficients are assumed to be different for each user, i.e., independent, while the large-scale ones are the same for all the $M$ antennas but depend on the user's position. Then, the channel matrix experienced by all the $K$ users can be expressed as
\begin{equation}\label{eq:chan_matrix}
\textbf{H} = \left( \begin{array}{ccc}
h_{1,1} & \cdots & h_{K,1} \\
\vdots & \ddots & \vdots \\
h_{1,M} & \cdots & h_{K,M} \end{array} \right) = \textbf{G}\textbf{D}^{1/2},
\end{equation}
where 
\begin{equation}\label{eq:small_scale}
\textbf{G} = \left( \begin{array}{ccc}
g_{1,1} & \cdots & g_{K,1} \\
\vdots &  \ddots & \vdots \\
g_{1,M} & \cdots & g_{K,M} \end{array} \right),
\end{equation}
\begin{equation}\label{eq:large_scale}
\textbf{D} = \left( \begin{array}{ccc}
d_{1} & & \\
&  \ddots &  \\
&  & d_{K} \end{array} \right),
\end{equation}

For uplink signal transmission, the received signal vector at a single BS, which we denote by $\textbf{y} \in \mathbb{C}^{M \times 1}$ has the following expression:
\begin{equation}\label{eq:uplink_rec_signal}
\textbf{y} = \sqrt{ \rho} \ \textbf{H} \ \textbf{x} + \textbf{n},
\end{equation}
where $\textbf{x} \in \mathbb{C}^{K \times 1}$ is the signal vector being transmitted by all users, i.e., transmit antennas, $\textbf{H} \in \mathbb{C}^{M \times K}$ is the uplink channel matrix defined in Eq. \ref{eq:chan_matrix}, $\textbf{n} \in \mathbb{C}^{M \times 1}$ is a zero-mean noise vector with complex Gaussian distribution and identity covariance matrix, and $\rho$ is the uplink transmit power. 

In this paper we consider OFDM block-based transmissions where the frequency-domain data symbols are randomly and independently drawn from a QPSK alphabet with normalized average energy. The transmitted sample from the $k$-th user, $x_{k}$, is the $k$-th element of $\textbf{x} =  [x_{1}, . . ., x_{K}]^{T} $ with $E[ | x_{k} | ^{2}] = 1 $, i.e., the OFDM symbols are normalized so that they present unit variance. 

As the small-scale fading coefficients for different users are assumed i.i.d. random variables with zero mean and unitary variance, the column channel vector from different users become asymptotically orthogonal as the number of receive antennas at the BS, $M$, grows to infinity \cite{marzetta:noncooperative}. Then
\begin{equation}\label{eq:vanishing_of_small_scale_fading}
\textbf{H}^{\text{H}} \textbf{H} = \textbf{D}^{1/2} \ \textbf{G}^{\text{H}} \textbf{D}^{1/2} \approx M  \textbf{D}^{1/2}  \textbf{I}_{K}  \textbf{D}^{1/2}  = M \textbf{D}  
\end{equation}
where $(\cdot)^{\text{H}}$ denotes the transpose-conjugate (Hermitian) operation. The conclusion shown in (\ref{eq:vanishing_of_small_scale_fading}) means we have $\textit{favorable propagation}$. Channel measurements presented in \cite{rusek:scaling_up_mimo} show that massive MU-MIMO systems have characteristics that approximate the favorable-propagation assumption fairly well, and therefore providing experimental justification for the assumption made above.

Based on the assumption of $\textit{favorable propagation}$ made above, and that the BS has perfect knowledge of the channel matrix, the achievable sum-rate, i.e., the total throughput, the channel, $\textbf{H}$, can offer is
\begin{equation}\label{eq:sum_rate}
\begin{split}
C = \text{log}_{2} \ \text{det}(\textbf{I} + \rho \textbf{H}^{\text{H}}\textbf{H}) \\ = \sum_{k=1}^{K} {\text{log}_{2}(1 + \rho M  d_{k})}  \ \   \frac{\frac{\text{bits}}{\text{s}}}{\text{Hz}},
\end{split}
\end{equation}

This is the mutual information between the input and the output of the Massive MIMO channel \cite{rusek:scaling_up_mimo}.

As said before, the channel vectors are asymptotically orthogonal when the number of antennas at the BS tends to infinity. Once $\textbf{D}$ is a diagonal matrix, the MF processing separates the signals from different users into different streams and there is asymptotically no inter-user interference \cite{marzetta:noncooperative}. Therefore, the transmission from each user can be seen as the transmission of an user through a SISO channel. This implies that MF detection at the BS is optimal when the number of receive antennas, $M$, is much larger then the number of transmit antennas and tends to infinity, i.e., $M \gg K$ and $M \to \infty$.

Throughout the paper we assume flat Rayleigh fading and perfect channel estimation at the BS. The assumption of flat Rayleigh fading means that the elements $h_{k,m}$ of the $M \times K$ channel $\textbf{H}$ are the complex channel gains from the transmit antennas to the receive ones.

\section{Uplink Detection}

\begin{figure*}
  \centering
  \begin{tabular}{ccc}
    \subfigure[ $10 \times 50$ \label{fig:50x10}]{\includegraphics[scale=0.42]{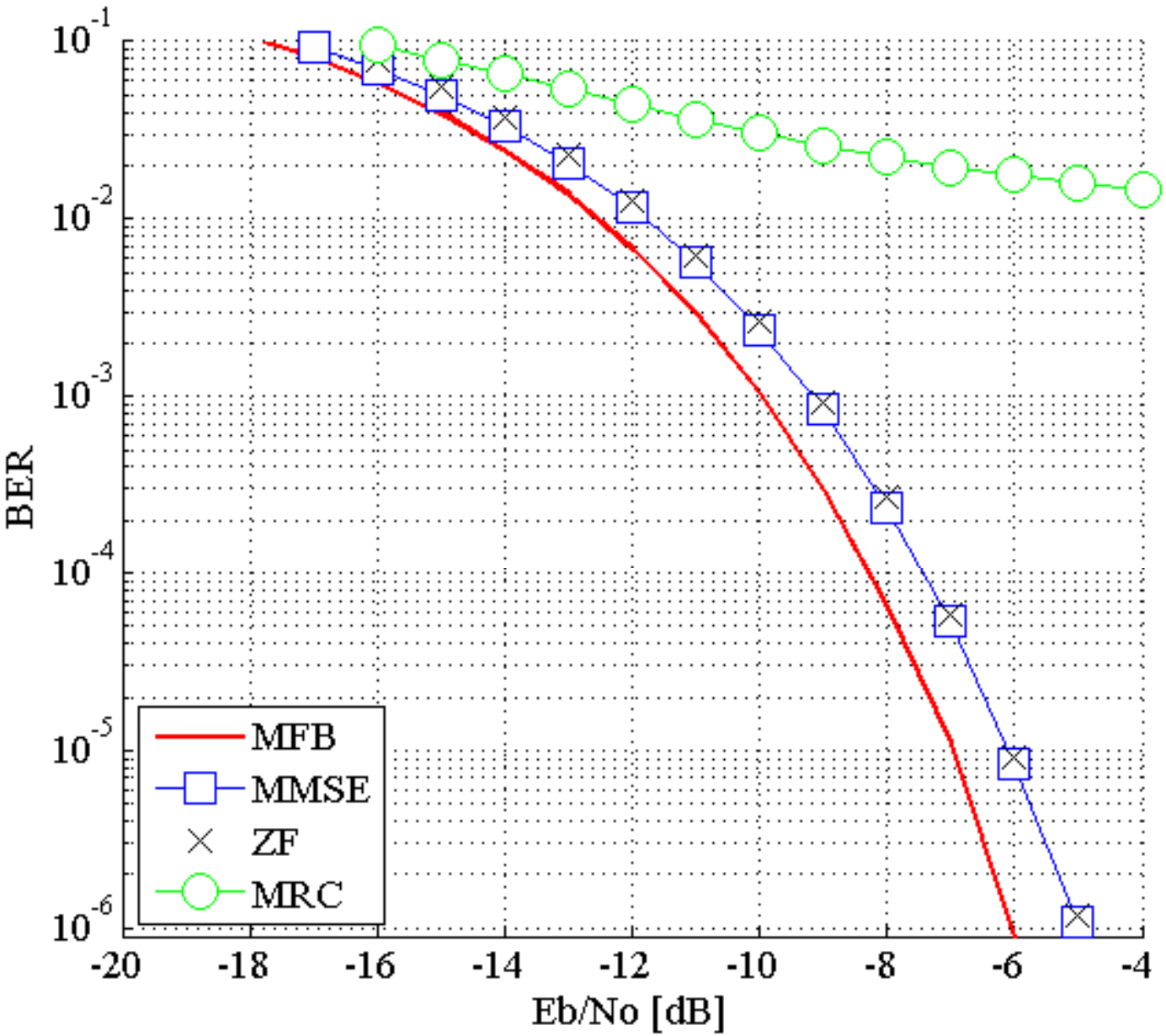}}\quad
    \subfigure[ $10 \times 100$ \label{fig:100x10}]{\includegraphics[scale=0.42]{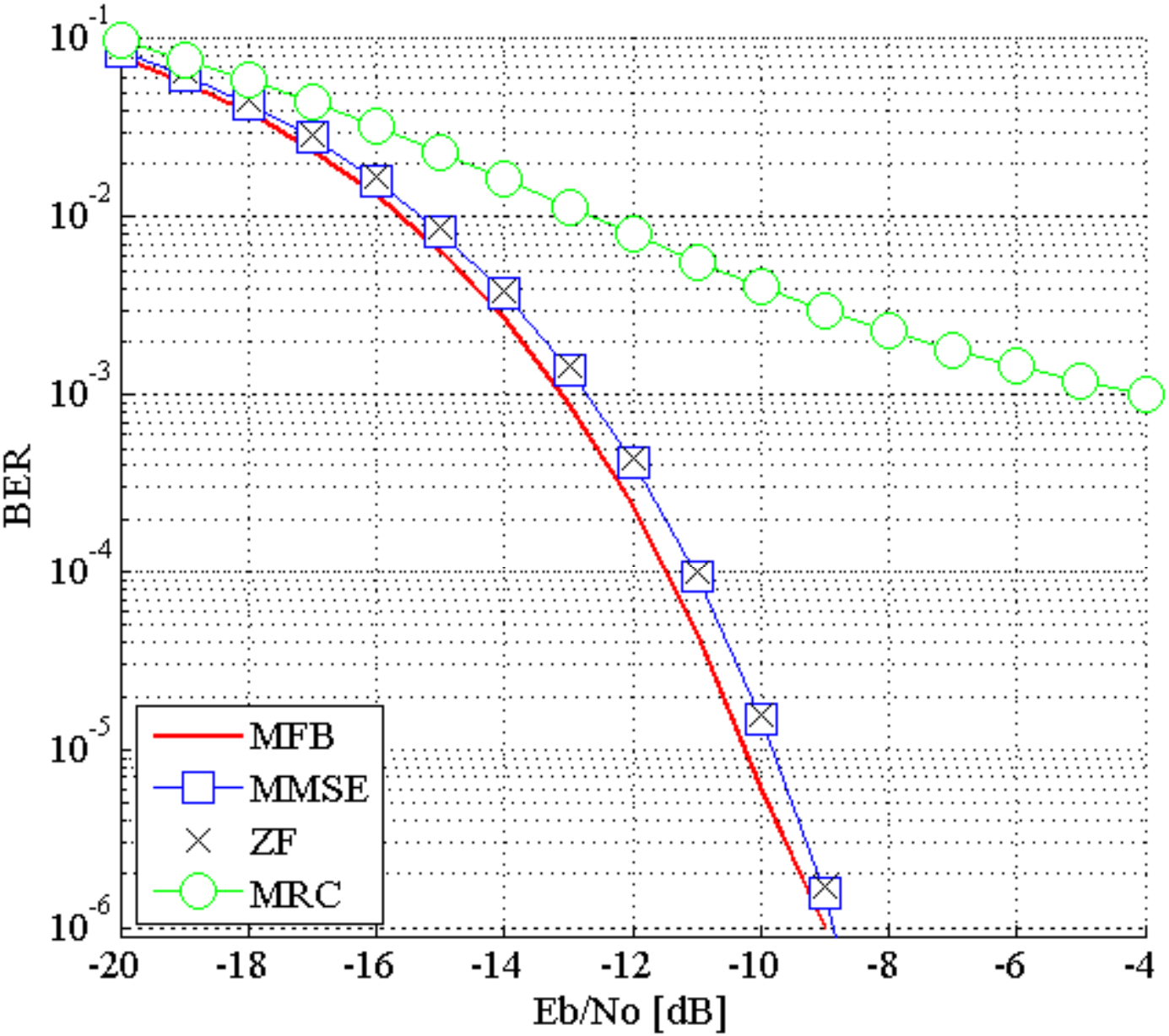}}\quad
    \subfigure[ $10 \times 250$ \label{fig:250x10}]{\includegraphics[scale=0.42]{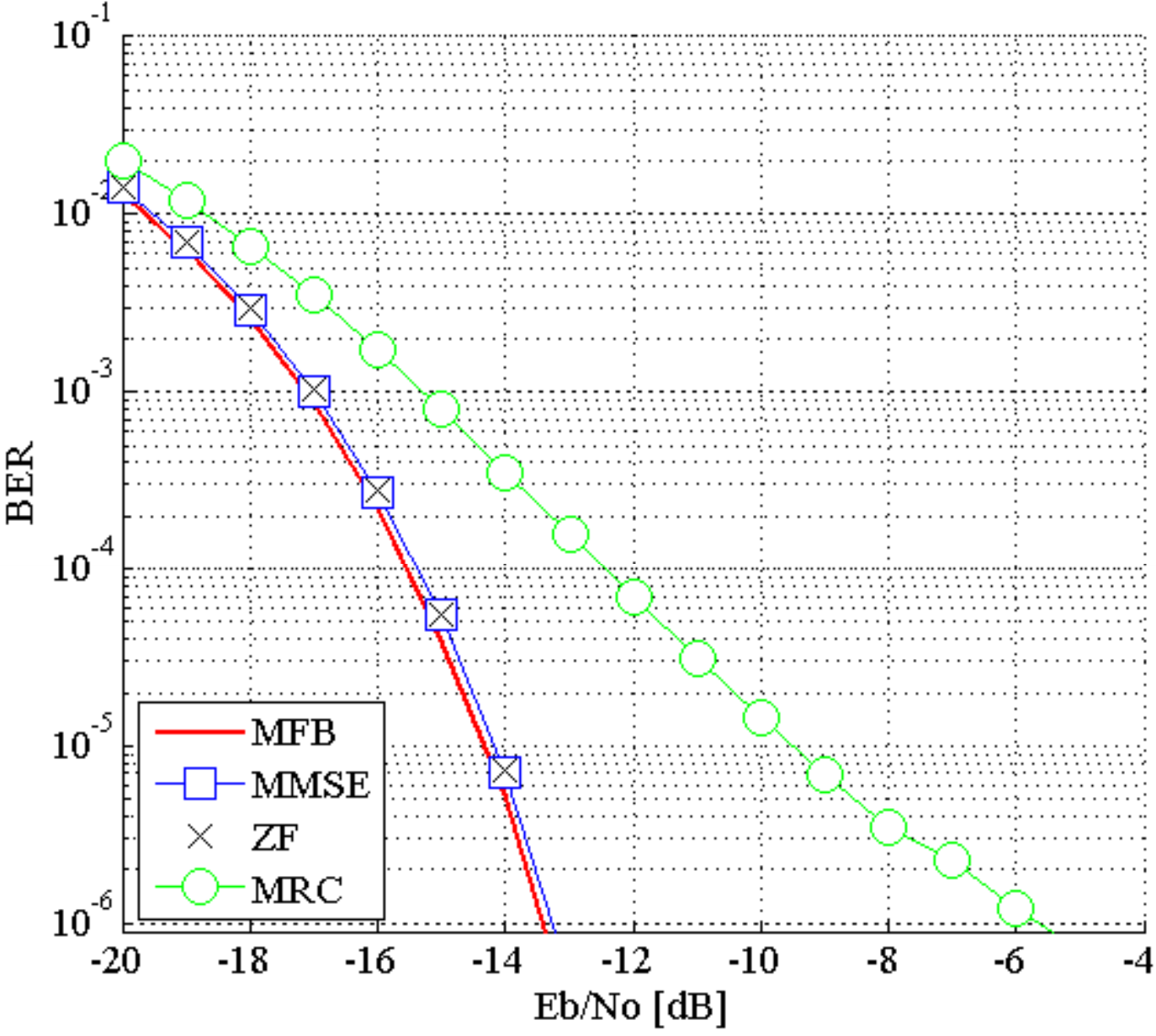}} \\
    
        \subfigure[ $10 \times 400$ \label{fig:50x10}]{\includegraphics[scale=0.42]{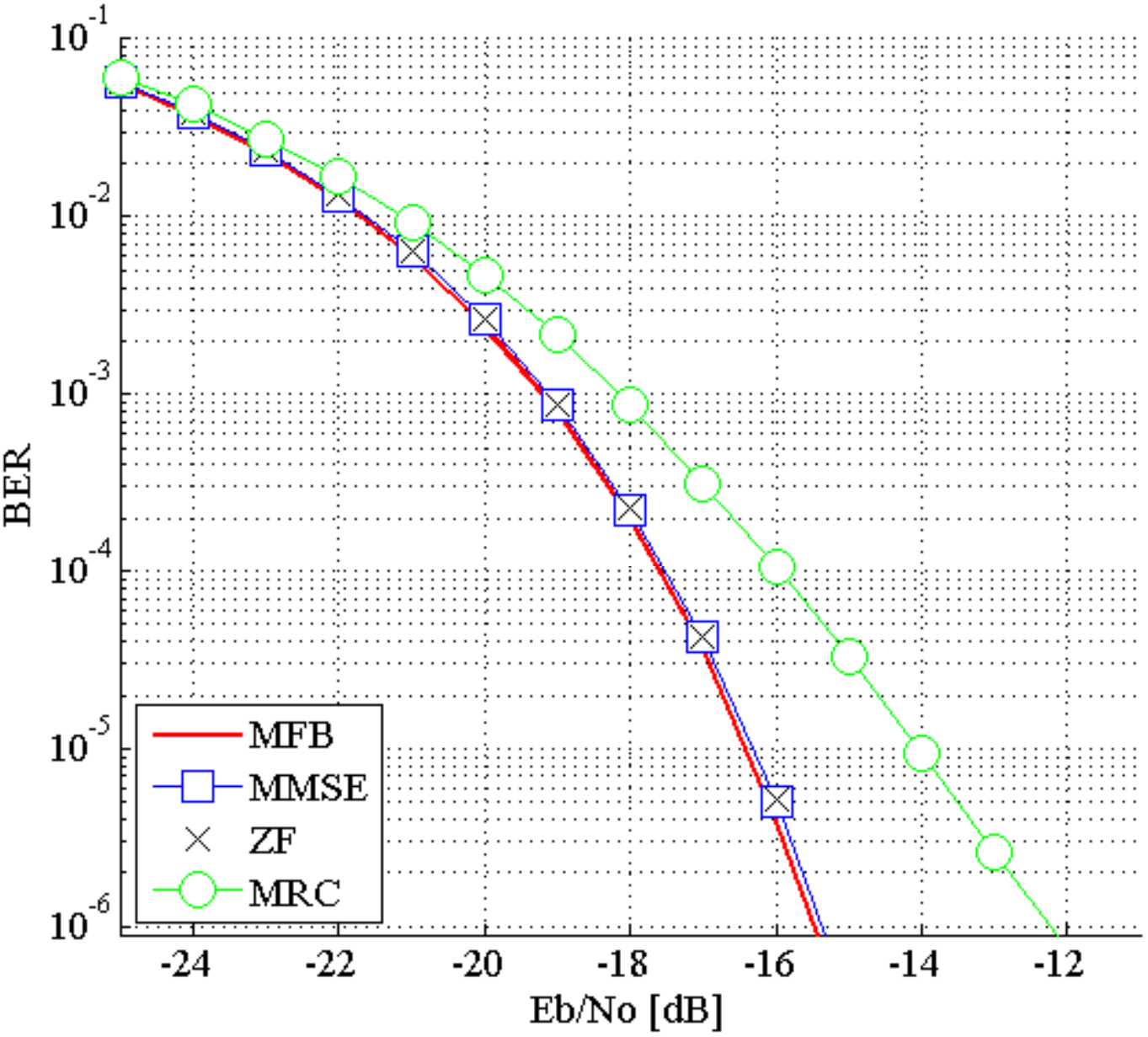}}\quad
    \subfigure[ $10 \times 500$ \label{fig:100x10}]{\includegraphics[scale=0.42]{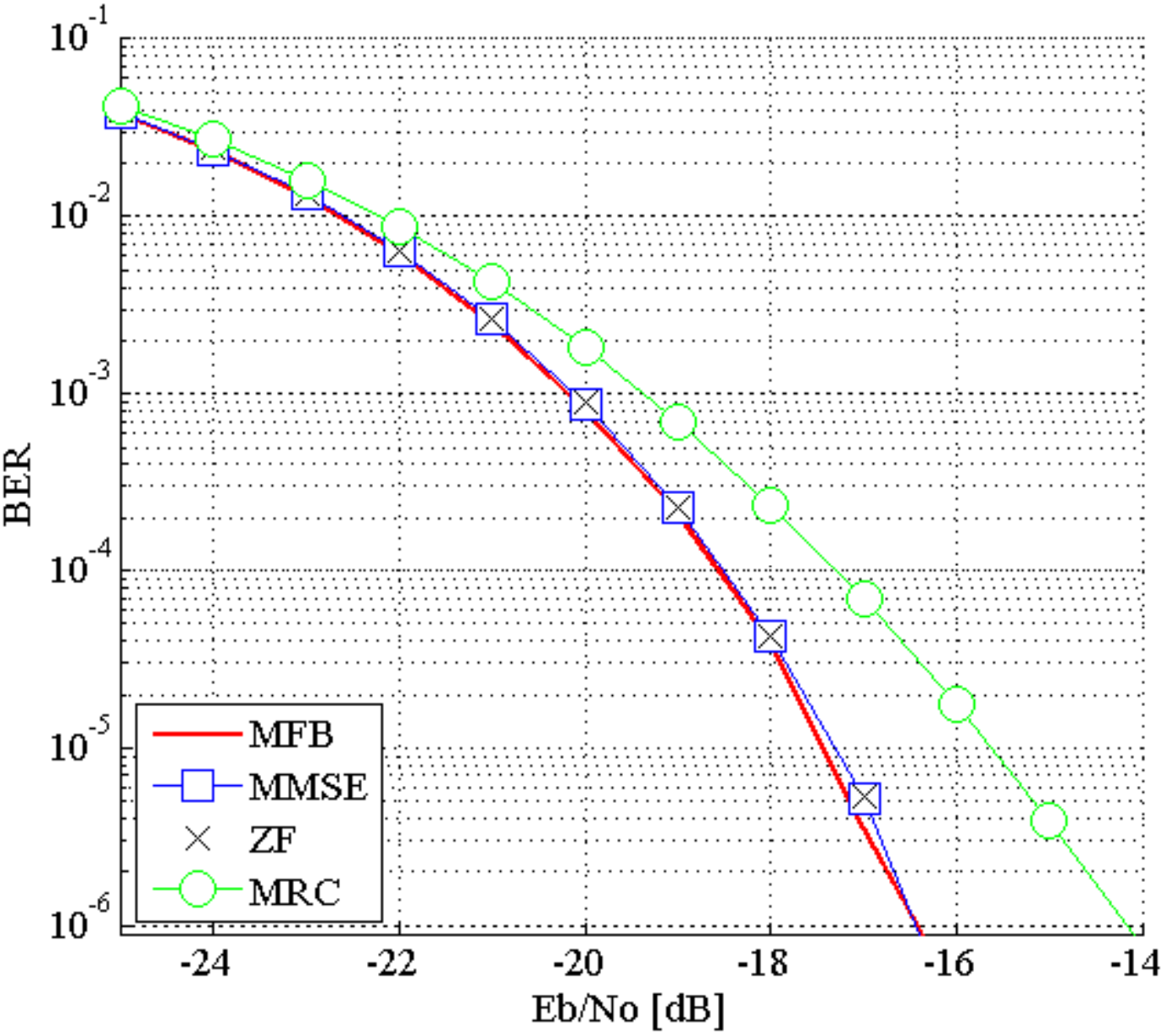}}
    
  \end{tabular}
  \caption{BER performance for Massive MU-MIMO detection on Uplink with $K=10$ and $M = 50 \ \text{(a)}, 100 \ \text{(b)}, 250 \ \text{(c)}, 400 \ \text{(d)} \ \text{and} \ 500 \ \text{(e)}$.}
  \label{fig:uplink_detection1}
\end{figure*}

Detection techniques must be employed in order to separate each of the data streams transmitted by the various users in a massive MU-MIMO system. 

Despite being the optimum detector, the maximum likelihood (ML) detector is highly complex. It has a complexity cost that grows exponentially with the number of transmit antennas, $K$, and modulation order used, which makes it very complex to be implemented in systems with hundreds of antennas as is the case with massive MIMO systems. Thus, when ML detection is impractical, we must resort to sub-optimal alternatives with reduced complexity \cite{barry:book}.

Linear sub-optimal detectors with low complexity, such as MRC (also known as MF), ZF and MMSE detectors, are feasible candidates for the role in massive MIMO systems. In the case where $M$ is much larger than $K$, i.e., $1 \ll K \ll M$, it is known that those linear detectors (MRC, ZF and MMSE) perform fairly well. They asymptotically achieve capacity as $M$ grows \cite{marzetta:noncooperative} and therefore we will only consider those detectors in this paper. 

We consider here the case where the BS has perfect CSI, i.e., it knows $\textbf{H}$. Let $\textbf{A}$ be and $M \times K$ linear detector matrix which depends on the channel $\textbf{H}$. By using a linear detector, the received signal is separated into different data streams by multiplying it with $\textbf{A}^{\text{H}}$ as follows
\begin{equation}\label{eq:lin_detector}
\textbf{r} = \textbf{A}^{\text{H}} \textbf{y}.
\end{equation}
where $\mathbf{r}$ is a $K \times 1$ signal vector containing the data streams of the $K$ single-antenna devices.

As stated before, we consider three conventional linear detectors MRC, ZF and MMSE, i.e.,
\begin{equation}\label{eq:linear_detectors}
\textbf{A} = \left\{
\begin{array}{c l}      
    \textbf{H}  & \text{for MRC}\\
    \textbf{H}(\textbf{H}^{\text{H}} \textbf{H})^{-1}  & \text{for ZF}\\
    \textbf{H}(\textbf{H}^{\text{H}} \textbf{H} + \frac{\sigma_{n}^{2}}{\sigma_{x}^{2}} \textbf{I})^{-1} & \text{for MMSE}
\end{array}\right.
\end{equation}
where $\sigma_{x}^{2}$ and $\sigma_{n}^{2}$ are the signal and noise variances, i.e., power, respectively.

A ZF linear detector chooses the matrix $\textbf{A}$ so as to eliminate interference completely, regardless of noise enhancement. Specifically, a ZF linear detector chooses $\textbf{A}$ so that $\textbf{A}\textbf{H}=\textbf{I}$. A drawback of the ZF linear detector is its insistence on forcing the interference to zero, regardless of the interference strength. ZF detectors discard any desired signal energy that lies in the interference subspace. A better strategy is to choose $\textbf{A}$ so as to balance the lost signal energy with the increased interference. From this point of view, it is much better to accept some residual interference if it allows the detector to capture more of the desired signal energy \cite{barry:book}. 

The MMSE linear detector chooses $\textbf{A}$ so as to minimize the mean squared error $ e = E[\| \textbf{A}^{\text{H}} \textbf{y} - \textbf{x} \|^{2}]$ directly, without any additional zero-forcing constraint that $\textbf{A}\textbf{H}=\textbf{I}$. Differently from the ZF detector, which minimizes interference but fails to treat noise, and differently from the MRC detector, which minimizes noise but fails to treat interference, the MMSE detector achieves an optimal balance of noise enhancement and interference suppression \cite{verdu:book}.
\break\indent The time complexity, or computational complexity, of the ZF and MMSE linear detectors is $\mathcal{O}(MK + MK^{2} + K^{3})$ \cite{rusek:scaling_up_mimo}. For MRC, the dominant computation is the matrix ordering, which presents a time complexity of $\mathcal{O}(MK)$ multiplications. As expected, the MRC complexity is much less than the one for both ZF and MMSE detectors.
\break\indent Through the law of large numbers is possible to show that MRC, ZF and MMSE detectors achieve the same capacity, i.e., uplink data rate, because when $M$ grows large, $\textbf{A}^{\text{H}}\textbf{A}$ tends to $\textbf{D}$, and therefore, the ZF and MMSE matrices tend to that of the MRC detector \cite{Ngo:energy_in_mu_mimo}.
\break\indent By using a large number of BS antennas, we can scale down the transmit power proportionally to $1/M$. At the same time we increase the spectral efficiency $K$ times by simultaneously serving $K$ users in the same time-frequency resource \cite{larsson:mimo_for_next_gen}. 

\section{Simulation Results}

In the following, we present a set of performance results for uncoded QPSK/OFDM uplink block transmission, with $N = 2048$ and $N_{CP} = 128$, within a Massive MU-MIMO $K \times M$ flat Rayleigh fading channel. Each of the single paths linking a TX to a RX antenna is modeled as a one-tap FIR filter with a complex coefficient drawn from a zero-mean and unit variance Gaussian random process. Each of the single paths is assumed to be uncorrelated to the other paths. 

The following performance results were obtained by random generation of a large number of channel (flat Rayleigh fading plus AWGN) and QPSK/OFDM symbol realizations by means of Monte-Carlo simulations, which involves an error counting procedure.

Figure \ref{fig:uplink_detection1} shows BER performance results for $K = 10$ and several values of $M$. The performances of the MRC, ZF and MRC detectors are compared to that of the matched filter bound (MFB) detector. This bound is also known in the literature as the single user MF bound, or the perfect interference-cancellation bound. Simply stated, the MFB is the performance of the MF receiver for the $i$-th user in the absence of other interferes \cite{barry:book}.

When $M \gg K$, both the multi-user interference and the fading effects tend to disappear and consequently, the BER performance for the MU-MIMO $K \times M$ flat Rayleigh fading channel becomes very close to that of the MFB.

The sub-figures in Figure \ref{fig:uplink_detection1} clearly show that the performance penalty, which is inherent to the reduced complexity MRC detector (when compared to the MMSE detector) can be made quite small, by increasing $M$ significantly. The figures also show that, under highly increased $M$ values, even the reduced complexity MRC detector can approximate the MFB.

MMSE is always better than MRC or ZF, however, its performance is very close to that of the ZF detector. MMSE always performs the best across the entire Eb/No range. As can be noticed, in each sub-figure of Figure \ref{fig:uplink_detection1} the gap between the performance of MRC and that of ZF (or MMSE) is reduced as the number of receive antennas, $M$, increases. Also, the gap between ZF and MMSE, which is already very small for small values of $M$, tends to vanish as $M$ increases. The MRC, ZF and MMSE linear detectors approach the MFB performance as $M$ increases, however, the gap between the latter two detectors and MFB diminishes faster.

Even though the MRC complexity is the lowest among the detectors assessed here, it is clear that its performance is the poorest one and consequently its adoption should be avoided in favor of more advanced detectors such as ZF and MMSE ones.  
 
\section{Conclusions}

This paper assessed the uplink performance of a massive MU-MIMO system with OFDM transmission, when simple linear detection techniques are employed at the BS.

Massive MU-MIMO systems offer the opportunity of increasing the spectral efficiency. This is possible with simple linear detectors such as MMSE, ZF or MRC. Generally, MMSE and ZF detectors outperform MRC owing to its ability to cancel out interference within the cell.

The BER results, presented and discussed in Section IV, confirm the "massive MIMO" effects provided by a number of receive antennas much larger than the number of transmit antennas, even when the low-complexity MRC detector is employed.
 

\end{document}